\documentclass[preprintnumbers,nofootinbib,noshowpacs,eqsecnum,superscriptaddress,prd]{revtex4}

\usepackage{graphicx}
\usepackage{amsmath,amsthm,amssymb}
\usepackage{mathrsfs,bbm}
\usepackage{array,color}
\usepackage{slashed}
\usepackage{tabularx}

\makeatletter       % Macros for arabic section numbering

\renewcommand{\p@subsection}{}

\makeatother

\DeclareMathOperator{\sgn}{sgn}			%signum

\def\imagetop#1{\vtop{\null\hbox{#1}}} 	%getting the images in tabulars top aligned

\begin{document}
\title{Scattering amplitudes from a deconstruction of Feynman diagrams}

\author{M. Maniatis}
    \email[E-mail: ]{maniatis8@gmail.com}
    
\author{C. M. Reyes}
    \email[E-mail: ]{cmreyes@uc.cl}
    
\affiliation{Departamento de Ciencias B\'a{}sicas, 
UBB, Casilla 447, Chill\'a{}n, Chile.}

\begin{abstract}
We show how to apply the BCFW recursion relation to Feynman 
loop integrals with the help of the Feynman-tree theorem. 
We deconstruct in this way all Feynman diagrams in terms of 
on-shell subamplitudes. Every cut originating from the Feynman-tree theorem corresponds to an
integration over the phase space of an unobserved particle pair.
We argue that we can calculate scattering amplitudes alternatively by the construction of
on-shell and gauge-invariant subamplitudes.
\end{abstract}

\maketitle

%%%%%%%%%%%%%%%%%%%%%%%%%%%%%%%%%%%%%%%%%%%%%%%%%%%%%%%%%%%%%%%%%%%%%%%%%%%%%%%
%%%%%%%%%%%%%%%%%%%%%%%%%%%%%%%%%%%%%%%%%%%%%%%%%%%%%%%%%%%%%%%%%%%%%%%%%%%%%%%
\section{Introduction}

In recent years lots of effort has been spent on the calculation of scattering amplitudes
without the usual Feynman diagram approach; see the reviews  
\cite{Bern:2007dw, Britto:2010xq, Ellis:2011cr, Elvang:2013cua,Dixon:2013uaa}.
With an increasing perturbation order or with an increasing number of
external particles, the number of Feynman diagrams grows in general rapidly.
However, the final scattering amplitude typically collapses to a 
rather short expression. One example is the very short
Parke-Taylor formula \cite{Parke:1986gb} for the
tree-level scattering amplitude for an arbitrary number of external gluons.
A tremendous simplification arises from the analytic continuation
of the momenta of external particles in the BCFW recursion relation 
approach \cite{Britto:2004ap, Britto:2005fq}. In particular all internal lines
become on-shell and gauge invariance holds diagram by diagram. 
This means that unphysical degrees-of-freedom do not enter 
the calculation. The BCFW recursion relations
are valid in gauge theories \cite{ArkaniHamed:2008yf}, 
but are restricted to tree diagrams (like
the Parke-Taylor formula). There are attempts to 
generalize the BCFW recursion relation to Feynman loop diagrams; see for instance
\cite{Boels:2010nw}.

Considering loop diagrams, there is the well-known unitarity cut
relation which reduces the loop order of a Feynman diagram. 
However, the unitarity cut only gives a relation to the imaginary part of a loop diagram.
It has been shown that by a generalization of this idea
one-loop amplitudes can be constructed
by generalized unitarity cuts \cite{Bern:1997sc,Ellis:2007br,Giele:2008ve,Ellis:2008ir}.

The Feynman-tree theorem, introduced
by Richard Feynman in 1963 \cite{Feynman:1963ax, Feynman:FTT},
opens the loops recursively and is not limited to
one-loop Feynman diagrams. In each recursion
step a loop is cut in all possible ways from single cuts
up to $n$ cuts, with $n$ the number of propagators in the considered loop.
In each recursion step a loop diagram with $n$ propagators is
decomposed into $n^2-1$ diagrams with at least one order of
loop reduced. Typically not all of these diagrams contribute.
In this sense the Feynman-tree theorem gives a natural
decomposition of Feynman diagrams into generalized cut diagrams.
Recursively, we can open all loops of a Feynman diagram, that
is, we can deconstruct eventually Feynman loop diagrams in terms of tree diagrams.
There is some recent interest in the Feynman tree theorem; see for instance
\cite{Brandhuber:2005kd, Catani:2008xa, CaronHuot:2010zt, Bierenbaum:2010cy}.

Here we shall argue that the Feynman-tree theorem combined
with the BCFW recursion relation allows for a systematic
deconstruction of Feynman diagrams into a product of
on-shell subamplitudes. Every cut gives a phase space integration
over a pair of unobserved particles. 
Subsequent application of the BCFW recursion relation through
analytic continuation of the external momenta allows for
a complete factorization of each tree diagram into simple vertex amplitudes.
In this way, any Feynman loop  diagram
is deconstructable in terms of on-shell subamplitudes. In particular,
each subamplitude is gauge invariant. 
One subtle point is that the Feynman-tree-theorem cuts give new external
particles with opposite momentum. Singularities are encoded
in the corresponding phase space integrations and we have to regularize
all diagrams to keep track of the singularities.
The key is to regularize the infrared and ultraviolet singularities
consistently. We will show that this is possible in dimensional 
regularization. 
The method of first opening the loops with the help of the Feynman-tree theorem
and then applying BCFW recursion
has been discussed in \cite{Maniatis:2015kex} with an explicit example of a two-point function 
employing Pauli-Villars regularization. 

Eventually we propose an alternative way to compose scattering 
amplitudes from a product of on-shell, gauge invariant subamplitudes.
Depending on the perturbation order considered, 
an appropriate number of unobserved particles has to be introduced.
Over the phase space of these unobserved particles has to
be integrated. In this way, no virtual particles appear and
we get the scattering amplitudes in a rather direct way.
The main focus in this work does not lie on the simplification of the
actual computation of amplitudes, since we trade loop integrations off against
phase-space integrations. This is in particular true
because of the rather large number of subamplitudes.
The point is that the composition of gauge-invariant
subamplitudes with on-shell particles represents 
physical scattering amplitudes, since 
unphysical off-shell degrees as well as gauge degrees are
systematically avoided. 

Since we dimensionally regularize infrared and ultraviolet singularities,
the Weyl spinor formalism can not be applied directly as long as the Weyl spinors
are given in two dimensions, corresponding to four-dimensional Dirac spinors.
It would be interesting to extend the method shown
here to the Weyl spinor formalism.  The integrations over the phase space 
could be performed in terms of Weyl spinors. Some work has
been done in context of Weyl spinors with generalized cuts; see 
for instance \cite{Anastasiou:2006gt,Mastrolia:2009dr,Britto:2010um}.

In the following section \ref{deconstruct} we present the details
of the argument to deconstruct
Feynman diagrams in terms of on-shell
gauge-invariant amplitudes.
In section \ref{vertex} we illustrate the method in an explicit example,
the electron-photon vertex correction.

\section{Deconstructing Feynman diagrams}
\label{deconstruct}

In order to make this paper more self contained we 
briefly remind the reader about the Feynman-tree theorem \cite{Feynman:1963ax, Feynman:FTT}.
A Feynman diagram with   
$l$ loops is reduced to diagrams with at most $l-1$ loops in each recursion step, that 
is, any loop Feynman diagram is eventually given in terms of tree diagrams.
This reduction follows from the introduction of 
{\em advanced propagators} $G_A(p)$ in addition to the usual 
propagators $G_F(p)$:
\begin{equation} \label{advanced}
G_F(p)  = \frac{i}{p^2 - m^2 + i \epsilon}, \qquad
G_A(p)  = \frac{i}{p^2 - m^2 - i \epsilon \sgn(p_0)}.
\end{equation}
With help of the identity
\begin{equation}
\frac{1}{x\pm i \epsilon} = P.V. \bigg(\frac{1}{x}\bigg) \mp i \pi \delta(x),
\end{equation}
with $P.V.$ the principal value prescription
we have 
\begin{equation} \label{feynmanadvanced}
G_A(p) = G_F(p) - 2 \pi \;\delta^{(+)}(p^2 - m^2)
\end{equation}
with $\delta^{(+)}(p^2 - m^2) = \theta(p_0)\delta(p^2 - m^2)$, as usual.
In an arbitrary loop diagram we subsequently consider each loop and replace the
usual Feynman propagators $G_F(p)$ by
the advanced propagators $G_A(p)$.
In the loop integration, owing to the advanced propagators, the poles of the
zero component are now above the real axis. 
We therefore see that the loop integral vanishes when we close the integration contour
on the lower half plane. With \eqref{feynmanadvanced}
we get
\begin{equation} \label{FeynmanTT}
\begin{split}
0 =& \int \frac{d^4 q}{(2\pi)^4}  N(q)\; \prod_i G_A^{(i)}(q-p_1-\ldots-p_i)\\ 
=& 
\int \frac{d^4 q}{(2\pi)^4}  N(q) \; \prod_i \bigg\{ G_F^{(i)}(q-p_1-\ldots-p_i) -
2 \pi \; \delta^{(+)}((k-p_1-\ldots-p_i)^2 -m^2) \bigg\}.
\end{split}
\end{equation}
The numerator of the loop diagram $N(q)$ depends in general also on the loop momentum,
as indicated.
We recognize the recursion relation in the last expression of \eqref{FeynmanTT}:
the original Feynman loop diagram is given in terms of diagrams with
propagators replaced by the delta distribution terms. 
In the product expansion we find that at least one propagator is cut reducing 
the number of loops about at least one unit.
Recursive application opens all loops in terms of
tree diagrams. 
Let us note that a loop diagram with $n$ propagators results in each 
recursion step to $n^2-1$ loop-reduced diagrams.

We observe that by the BCFW recursion relations \cite{Britto:2004ap,Britto:2005fq} we can 
express the tree amplitudes, resulting from the Feynman-tree theorem, in terms of on-shell amplitudes. 
The basic idea of the BCFW recursion relations 
is analytic continuation of the external momenta. In this way 
tree amplitudes factorize into on-shell subamplitudes without violation of
momentum conservation.
In an arbitrary tree amplitude let us denote the $e$ external momenta
by $p^\mu_i$ with $i=1,\ldots,e$.
These external momenta are shifted,
\begin{equation} \label{shift}
\hat{p}_i^\mu = p_i^\mu + z \cdot r_i^\mu
\end{equation}
with one common $z \in \mathbb{C}$ and appropriately chosen vectors $r_i$,
such that $r_1+ \ldots + r_e=0$, $r_i\cdot r_j=0$ ($i,j \in \{1,\ldots,e\}$), $r_1 p_1=...=r_e p_e  =0$, 
keeping in particular
$\hat{p}_i^2= p_i^2$ invariant. 
By the analytic continuation \eqref{shift} a tree amplitude $A$ can be decomposed in
terms of on-shell subamplitudes,
\begin{equation} \label{BCFW}
A = - \sum_{z_I} \text{Res}_{z=z_I} \frac{\hat{A}(z)}{z} + B =
\sum_{\text{diagram }I} \hat{A}_L(z_I) \cdot \frac{1}{P_I^2} \cdot \hat{A}_R(z_I) + B.
\end{equation}
Here, $\hat{A}(z)$ denotes the shifted amplitude with the position of the poles in the complex
plane at $z=z_I$. On the right-hand side we
have the on-shell subamplitudes $\hat{A}_L(z_I)$ and $\hat{A}_R(z_I)$, with a propagator factor $1/P_I^2$.
The term $B$ denotes the residues of the poles of $\hat{A}(z)$ at $|z| \to \infty$. 
 In case of a vanishing term $B$ we have on the right-hand side
the desired factorization into on-shell subamplitudes.
It has been shown that the term $B$ vanishes for an appropriate shift of the external 
momenta in gauge theories \cite{ArkaniHamed:2008yf}. 
An example of a theory, where we do 
not have a vanishing contribution $B$ is $\phi^4$ theory, as discussed in \cite{Feng:2009ei}.
Considering gauge theories we have therefore the appropriate method we can apply
to the diagrams originating from the cut diagrams, which themselves come from the
Feynman-tree theorem. Note that all internal particles become on-shell, on the one hand
from the cuts of the Feynman-tree theorem and on the other hand from the BCFW recursion
relations. 
Every cut corresponds to a pair of unobserved particles. These pairs have opposite momenta
and correspond to an unobserved particle-antiparticle state. We have to sum over
all possible spins (and other degrees-of-freedom) of the unobserved particles. 

We observe that we equivalently can start the construction of scattering amplitudes
by gauge invariant on-shell diagrams: to a given perturbation order we have to
consider an appropriate number of unobserved particles. In particular all particles are
on-shell and each contribution is gauge invariant. All unphysical off-shell degrees-of-freedom
as well as gauge degrees do not enter the calculation. 
In the next section we shall demonstrate this in an explicit example, the
electron-photon vertex correction.

\section{Example: electron-photon vertex correction }
\label{vertex}

\begin{figure}[Ht!]
\includegraphics[width=0.2\textwidth]{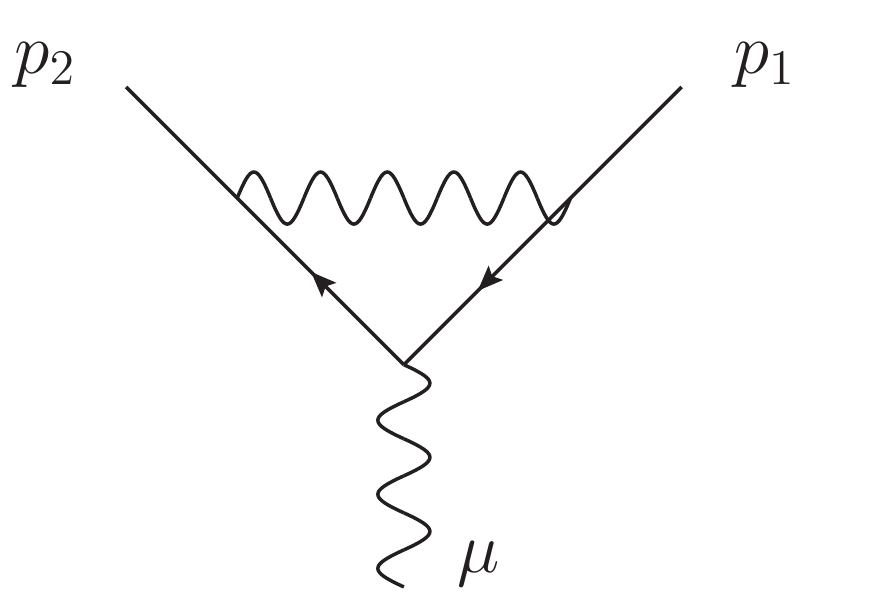}
\caption{\label{gm2-crop}The one-loop Feynman diagram
of the vertex correction to the electron-photon scattering.
}
\end{figure}
We want to demonstrate the deconstruction of Feynman integrals
in a simple example, the vertex correction to the scattering
of an electron with a photon.
The Feynman diagram of the vertex correction is shown in Fig. \ref{gm2-crop}.
We have to integrate over the undetermined loop momentum $q$ and find in
the usual Feynman diagram approach
\begin{equation} \label{amp1loop}
A_3(p_1;s_1,p_2;s_2,p_3;\lambda) = 
\bar{u}(p_2,s_2)\cdot \Gamma_\mu(p_1, p_2)\cdot u(p_1,s_1) \epsilon^\mu(p_3, \lambda)
\end{equation}
with 
\begin{equation} \label{amp1Gamma}
\Gamma_\mu(p_1, p_2) = 
  e^3 \mu^{4-D} 
 \int \frac{ d^D q}{(2\pi)^D}
 \frac{\gamma_\alpha (\slashed{q}+\slashed{p}_2+m) \gamma_\mu (\slashed{q}+\slashed{p}_1+m) \gamma^\alpha}
 {q^2 [(q+p_1)^2-m^2] [(q+p_2)^2-m^2]}.
\end{equation}
The momenta of the external particles are denoted by $p_1$, $p_2$, $p_3$, the mass of 
the electron by $m$ and the spins of the 
electrons by $s_1$ and $s_2$, whereas the polarization of the photon is denoted by $\lambda$.
Since the diagram is singular we employ dimensional regularization which gives
the usual $D$-dimensional integral measure together with an arbitrary mass scale $\mu$.
Using Passarino-Veltman tensor reduction \cite{Passarino:1978jh} and standard scalar integrals we find the well-known
result
\begin{multline}
\Gamma_\mu(p_1, p_2) = 
\frac{i e^3}{(4\pi)^2}\bigg\{ 
\gamma_\mu \big(
4 B_0(m^2,m,0) - 3 B_0((p_1-p_2)^2,m,m)
+ (4 m^2 -2 (p_1-p_2)^2) C_0(m^2,m^2,(p_1-p_2)^2,m,m,0) - 2
\big)\\
+ (p_1+p_2)_\mu \frac{2 m}{4m^2-(p_1-p_2)^2}
\big(
B_0((p_1-p_2)^2,m,m) - 2 B_0(m^2,m,0) + B_0(0,0,m) + 1 
\big) \bigg\}
\end{multline}
with $B_0$ and $C_0$ the standard scalar integrals (see for instance
\cite{Denner:1991kt} for explicit expressions for the scalar integrals).
%
%\begin{equation}
%\begin{split}
%&B_0(0,0,m) = \Delta - \ln \big(\frac{m^2}{\mu^2}) + 1,\\
%&B_0(m^2,m,0) = \Delta - \ln \big(\frac{m^2}{\mu^2}) + 2,\\
%&B_0(s,m,m) = \Delta - \ln \big(\frac{m^2}{\mu^2}) + \frac{s}{6 m^2} + ?,\\
%\end{split}
%\end{equation}
In particular the ultraviolet and infrared singularities
of this vertex correction are encoded in these scalar integrals.

\begin{figure}
\begin{tabular}[t]{llllll}
1 & 
\imagetop{\includegraphics[width=0.15\textwidth]{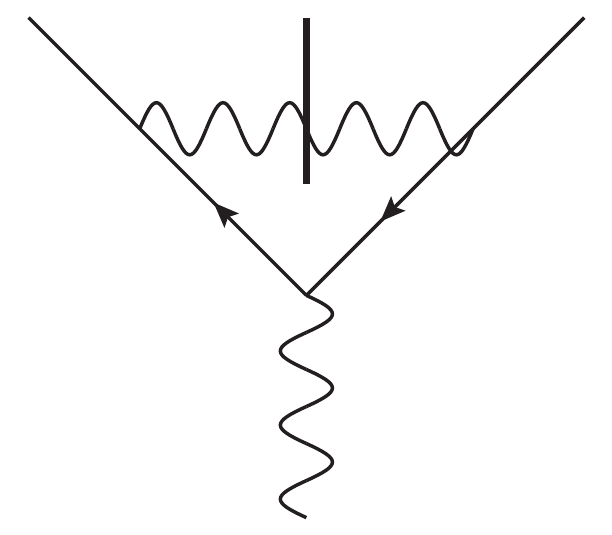}}\qquad\qquad
&
2 &
\imagetop{\includegraphics[width=0.15\textwidth]{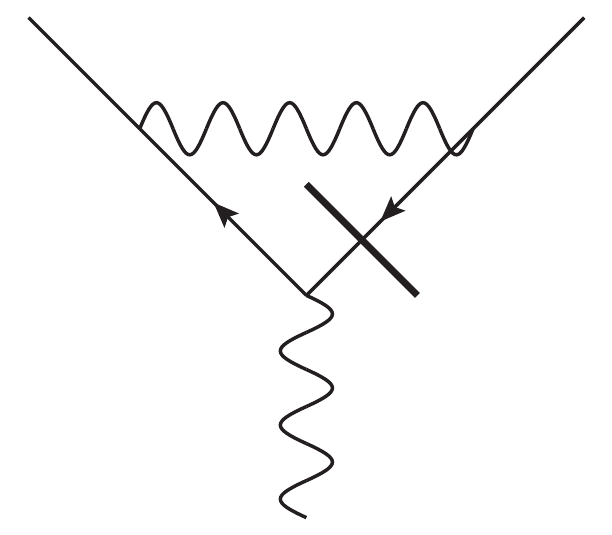}}\qquad\qquad
&
3 &
\imagetop{\includegraphics[width=0.15\textwidth]{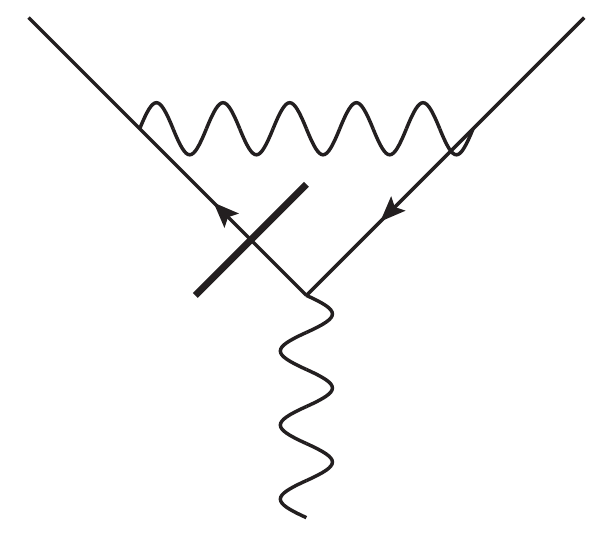}}\qquad\qquad
\\
4 & 
\imagetop{\includegraphics[width=0.15\textwidth]{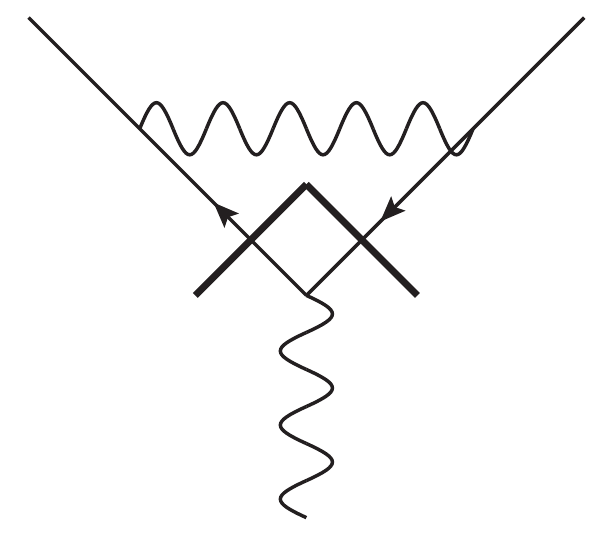}}\qquad\qquad
&
5 &
\imagetop{\includegraphics[width=0.15\textwidth]{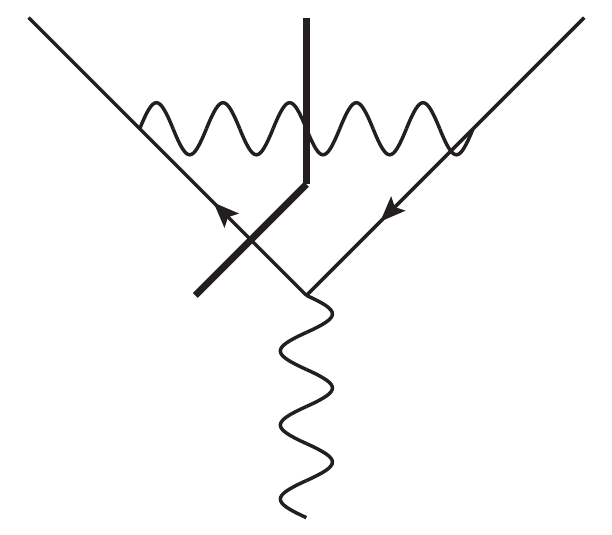}}\qquad\qquad
&
6 &
\imagetop{\includegraphics[width=0.15\textwidth]{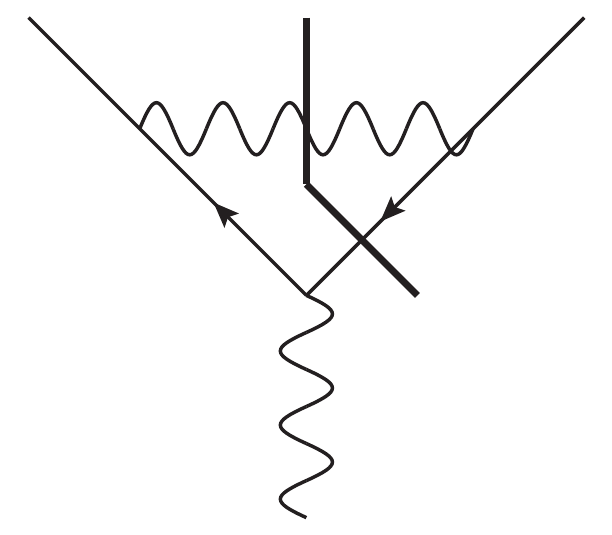}}\qquad\qquad
\\
7 &
\imagetop{\includegraphics[width=0.15\textwidth]{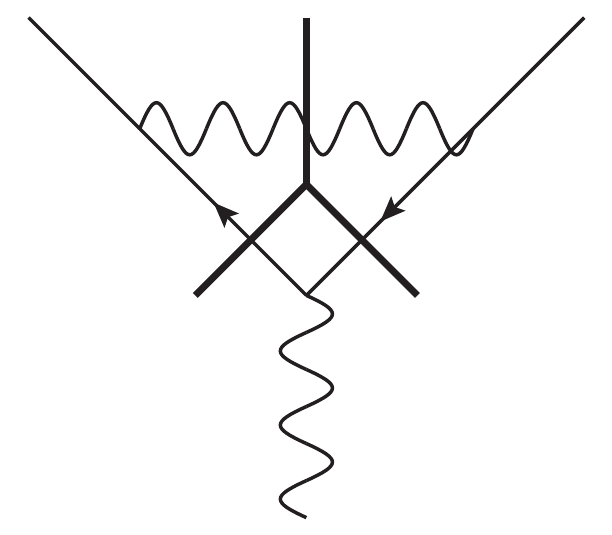}}\qquad\qquad
\end{tabular}
\caption{\label{ampFTT}The cut diagrams obtained from the one-loop vertex 
correction by the application of the Feynman-tree theorem.}
\end{figure}

We start the deconstruction of the vertex correction \eqref{amp1loop}, \eqref{amp1Gamma}
with the Feynman-tree theorem \eqref{FeynmanTT}. Since the loop
has three internal lines, there are $2^3-1=7$ cut diagrams
as shown in Fig. \ref{ampFTT}. Moreover, the loop diagram is given in 
$D$ dimensions, so the integration measure in \eqref{FeynmanTT} has to be generalized 
to $D$ dimensions.
The first row in Fig. \ref{ampFTT} shows the single cuts, the second row the double cuts,
and the last row the triple cut. 
The separate cut diagrams are 
\begin{equation}
A_3^{(i)}(p_1;s_1,p_2;s_2,p_3;\lambda) = \bar{u}(p_2,s_2) \cdot \Gamma_{\mu}^{(i)} (p_1, p_2) \cdot u(p_1,s_1) \epsilon^\mu(p_3, \lambda) ,
\end{equation}
where $i$ denotes the diagram number as given in Fig. \ref{ampFTT}.

The first single-cut diagram with the cut in the photon propagator, with view on
\eqref{FeynmanTT}, reads 
\begin{equation} \label{Gamma1FTT}
\begin{split}
\Gamma_\mu^{(1)}(p_1, p_2) = &
  - e^3 \mu^{4-D}
 \int \frac{ d^D q}{(2\pi)^D} 2 \pi\cdot \delta^{(+)}(q^2)
 \frac{\gamma_\alpha (\slashed{q}+\slashed{p}_2+m) \gamma_\mu (\slashed{q}+\slashed{p}_1+m) \gamma^\alpha}
 {[(q+p_1)^2-m^2] [(q+p_2)^2-m^2]}\\
 = &
  e^3 \mu^{4-D} \sum_{\lambda_1}
  \int \frac{d^{D-1} k_1}{(2 \pi)^{D-1} 2 k_{10}}
 \frac{\gamma_\alpha (\slashed{k}_1+\slashed{p}_2+m) \gamma_\mu (\slashed{k}_1+\slashed{p}_1+m) \gamma_\beta}
 {[(k_1+p_1)^2-m^2] [(k_1+p_2)^2-m^2]} \cdot \epsilon^\alpha(k_1,\lambda_1) \epsilon^{\beta *}(k_1,\lambda_1) \\
% \frac{\gamma_\alpha (\slashed{q}+\slashed{p}_2+m) \gamma_\mu (\slashed{q}+\slashed{p}_1+m) \gamma^\alpha}
% {(2 q p_1) (2 q p_2)}.
 \end{split}
 \end{equation}
In the last step we have rewritten the loop integral measure
as a phase space integral over the unobserved on-shell photon pair and relabeled 
$q$ as $k_1$. We have to sum over the on-shell helicities of the photon pair.

For the second single-cut diagram we find
\begin{equation}
\Gamma_\mu^{(2)}(p_1, p_2) = 
  - e^3 \mu^{4-D} \sum_{s_3}
 \int \frac{d^{D-1} k_1}{(2 \pi)^{D-1} 2 k_{10}}
 \frac{\gamma_\alpha (\slashed{k}_1-\slashed{p}_1+\slashed{p}_2+m) \gamma_\mu u(k_1,s_3) \cdot \bar{u}(k_1,s_3) \gamma^\alpha}
  {(k_1-p_1)^2 [(k_1-p_1+p_2)^2-m^2]}.
 \end{equation}
We have to sum over the helicities 
of the unobserved on-shell Dirac electron. We see the explicit integration over
the phase space of the unobserved electron.

The calculation of the third single-cut amplitude can be performed analogously.
Let us proceed with the double-cut diagrams. 
We explicitly calculate the diagram 4,
\begin{equation}
\begin{split}
\Gamma_\mu^{(4)}(p_1, p_2) = &
 e^3 \mu^{4-D} 
 \int \frac{ d^D q}{(2\pi)^D}\;2\pi\;\delta^{(+)}((q+p_1)^2-m^2)\;2\pi\;\delta^{(+)}((q+p_2)^2-m^2)
 \cdot\\
 & \qquad \qquad
  \frac{\gamma_\alpha (\slashed{q}+\slashed{p}_2+m) \gamma_\mu (\slashed{q}+\slashed{p}_1+m) \gamma^\alpha}
% \frac{\gamma_\alpha u(q+p_2,s_2)\bar{u}(q+p_2,s_2)  \gamma_\mu  u(q+p_1,s_1)\bar{u}(q+p_1,s_1) \gamma^\alpha}
 {q^2}\\
  = & 
 e^3 \mu^{4-D} 
 \sum_{s_3, s_4}
 \int \frac{d^{D-1} k_1}{(2 \pi)^{D-1} 2 k_{10}} \frac{d^{D-1} k_2}{(2 \pi)^{D-1} 2 k_{20}}
 (2 \pi)^D \delta^{(D)}(p_1-p_2+k_2-k_1) \cdot \\
 & \qquad \qquad
 \frac{\gamma_\alpha u(k_2,s_4) \cdot \bar{u}(k_2,s_4)  \gamma_\mu 
 u(k_1,s_3) \cdot \bar{u}(k_1,s_3) \gamma^\alpha}
 {(k_1-p_1)^2}
\end{split}
\end{equation}
We see the explicit phase space integration over the two unobserved Dirac electrons.
The other two double-cut diagrams $\Gamma_\mu^{(5)}$ and $\Gamma_\mu^{(6)}$ can 
be performed analogously.
Let us eventually calculate the triple-cut diagram $\Gamma_\mu^{(7)}$.
From the three cuts in Fig. \ref{ampFTT} we see that this diagram has the product
of three delta distributions in the integrand, that is,
$ \delta^{(+)}(q^2) \cdot \delta^{(+)}((q+p_1)^2-m^2) \cdot \delta^{(+)}((q+p_2)^2-m^2)$.
These three delta distributions are only simultaneously fulfilled for $q p_1= q p_2 =0$ which
in general can not be arranged for the two
fixed external momenta $p_1$ and $p_2$. Therefore this triple-cut diagram vanishes in general.

\begin{figure}
\begin{tabular}[t]{llllll}
1 & 
\imagetop{\includegraphics[width=0.15\textwidth]{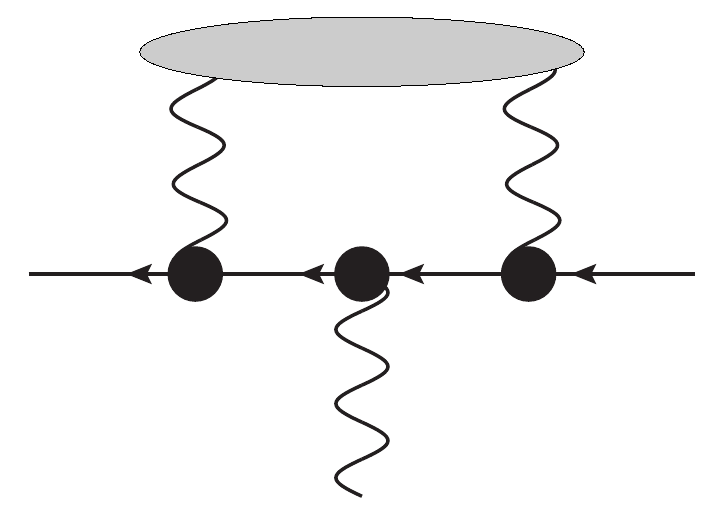}}\qquad\qquad
&
2 &
\imagetop{\includegraphics[width=0.15\textwidth]{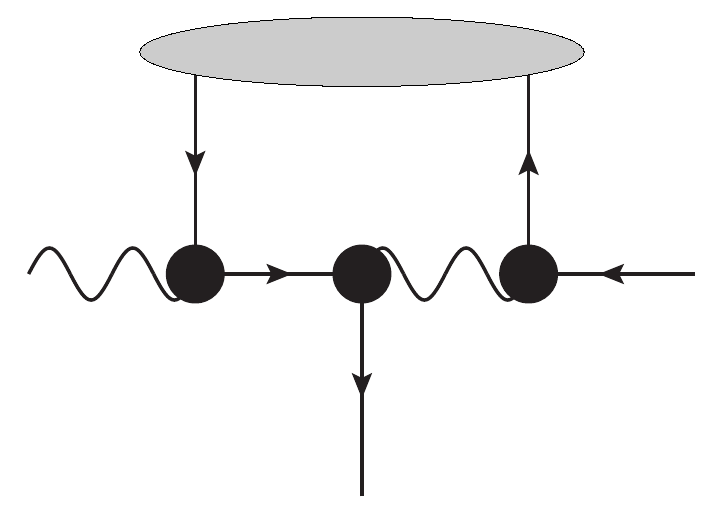}}\qquad\qquad
&
3 &
\imagetop{\includegraphics[width=0.15\textwidth]{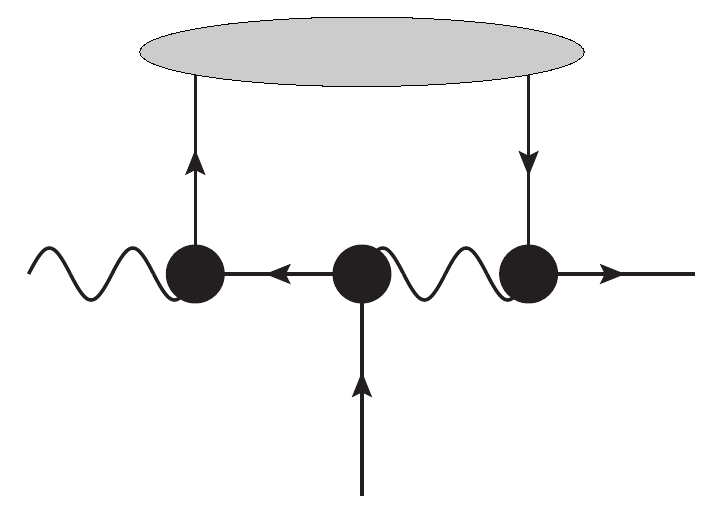}}\qquad\qquad
\\
4 & 
\imagetop{\includegraphics[width=0.15\textwidth]{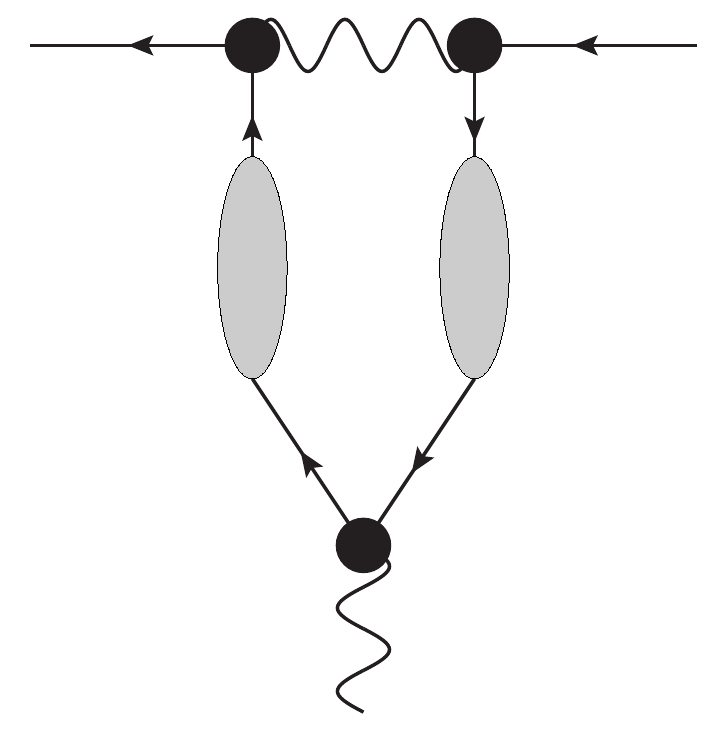}}\qquad\qquad
&
5 &
\imagetop{\includegraphics[width=0.15\textwidth]{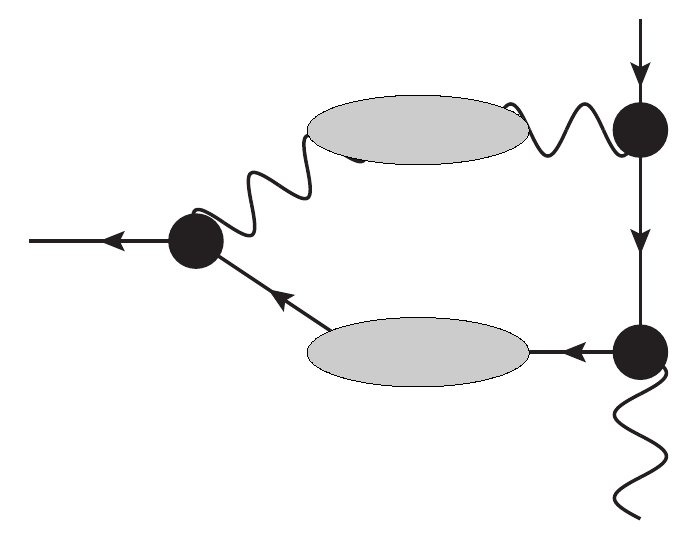}}\qquad\qquad
&
6 &
\imagetop{\includegraphics[width=0.15\textwidth]{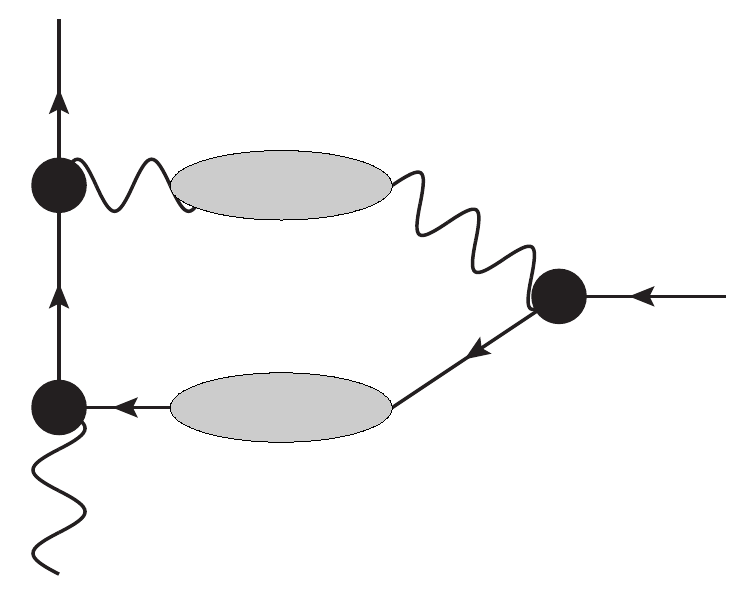}}\qquad\qquad
\\
7 &
\imagetop{\includegraphics[width=0.14\textwidth]{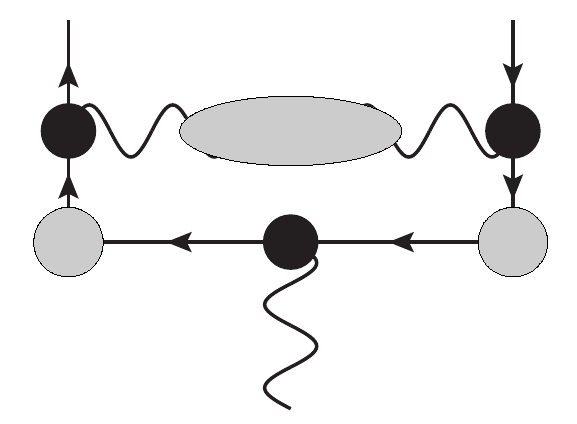}}\qquad\qquad
\end{tabular}
\caption{\label{ampBCFW}The contributions to the on-shell scattering amplitudes
of the electron-photon vertex correction. Full dots correspond to on-shell vertices
and the gray areas to the phase-space integration over the unobserved particle-antiparticle pair.}
\end{figure}

Having represented all diagrams as tree level diagrams we are in the position to
deconstruct them further by the application of the BCFW recursion relation.
In the cases of single cut diagrams (1-3), given in the first row in Fig. \ref{ampFTT} we 
apply two subsequent recursion steps corresponding to the two remaining propagators. 
In the cases of the double cut diagrams (4-6) we 
only have to apply once the BCFW recursion relation. We note that 
the BCFW recursion relation is valid in $D$ dimensions. In this way every diagram
can be eventually expressed in terms of on-shell scattering amplitudes.
As an example we calculate the 
diagram $\Gamma_\mu^{(1)}$ in Fig. \ref{ampFTT} explicitly in terms of the BCFW recursion relation.
This diagram can be computed by
analytic continuation of two external momenta; we choose to shift the two momenta
$p_1$ and $p_2$ in the form \eqref{shift}
\begin{equation} \label{shifting}
\hat{p}_1= p_1 + z r, \qquad
\hat{p}_2= p_2 + z r
\end{equation}
with $r$ chosen such that $r p_1 = r p_2 =r^2 =0$, keeping 
$\hat{p}_1^2= p_1^2$ and $\hat{p}_2^2= p_2^2$ invariant.
We start with the right-hand side of expression \eqref{Gamma1FTT}, that is,
with one cut, corresponding to one unobserved photon pair. 
Applying the shift \eqref{shifting} we get
\begin{multline} \label{Gamma1BCFW}
A_3^{(1)}(p_1;s_1,p_2;s_2,p_3;\lambda)=
e^3 \mu^{4-D} \sum_{s_3, s_4, \lambda_1}
  \int \frac{d^{D-1} k_1}{(2 \pi)^{D-1} 2 k_{10}} \quad \cdot \\
\bar{u}(\hat{p}_2, s_2) \slashed{\epsilon}(k_1,\lambda_1) u(k_1+\hat{p}_2, s_4)
\cdot
\frac{1}{P_2^2}
\cdot
\bar{u}(k_1+\hat{p}_2, s_4) \slashed{\epsilon}(p_3, \lambda) u(k_1+\hat{p}_1, s_3)
\cdot 
\frac{1}{P_1^2}
\cdot
\bar{u}(k_1+\hat{p}_1, s_3)
\slashed{\epsilon}^*(k_1, \lambda_1)
u(\hat{p}_1,s_1),
\end{multline}
with $P_1^2 = (k_1+p_1)^2-m^2$ and $P_2^2 = (k_1+p_2)^2-m^2$, following the
rules of BCFW.
We have with \eqref{Gamma1BCFW} the desired factorization into three
on-shell vertices.
The integration over the phase space and summation over the helicity 
of the unobserved photon originate
from the Feynman-tree theorem single cut. We have to sum over the
different spins and helicities of the internal lines. Let us note that
all the shifted momenta can be replaced by the external momenta of the amplitude.
The result \eqref{Gamma1BCFW} is depicted
in the diagram 1 of Fig. \ref{ampBCFW}. The full dots 
denote the on-shell vertices and the grey areas denote external 
but unobserved pairs of particles.
In an analogous way we can compute all the contributions to
the electron-photon vertex correction as shown in Fig. \ref{ampBCFW}.
We see that every cut corresponds to a pair of unobserved particles. 

This example shows that we can compute
the scattering amplitude starting with the amplitudes 
as presented in Fig. \ref{ampBCFW}. 
Since we are focussing on the vertex correction to order $e^3$ 
these are all contributions we have to consider to this order.
Obviously, in this way we get the scattering
amplitud by on-shell, gauge-invariant subamplitudes
and the elementary building block is the
fermion-photon vertex.

\section{Conclusions}

The BCFW recursion relations are a very elegant way to compute {\em tree} scattering
amplitudes in terms of gauge invariant on-shell subamplitudes.
We have shown that loop amplitudes can be decomposed by the BCFW 
recursion relations if the loops are recursively opened by
the application of the Feynman-tree theorem.
Since the cut diagrams of the Feynman-tree theorem
are singular in general, we have to consistently 
regularize these singularities. Here we used dimensional
regularization and used the fact that the BCFW recursion relations
are valid in $D$ dimensions. We note that the method is not
limited to a certain perturbation order. 

In an explicit example, namely, the electron-photon vertex correction, we
have shown the method in practice. We have seen that for every
cut, coming from the Feynman-tree theorem, we encounter an unobserved
particle pair. Eventually, in the deconstruction of loop diagrams,
all particles become on-shell, on the one hand from the cuts and on 
the other hand from the BCFW factorization. Moreover, every contribution to the scattering
amplitude is separately gauge invariant. 

We have seen that we can start the amplitude calculation in an alternative way
from gauge-invariant, on-shell subamplitudes. 
Systematically, unphysical degrees-of-freedom, that is,
off-shell modes and gauge degrees-of-freedom are avoided. In contrast, in
the usual Feynman diagram approach, gauge invariance is in general 
violated in each Feynman diagram and only restored in the sum of diagrams.

\section*{Acknowledgement}
We would like to thank 
Simon Caron-Huot and Otto Nachtmann
for many valuable comments and suggestions.
This work is supported partly by the 
Chilean research project FONDECYT, with project numbers
1140568, 1140781 as well as by the group of {\em F\'{i}sica
de Altas Energias} of the UBB, Chile.

%%%%%%%%%%%%%%%%%%%%%%%%%%%%%%%%%%%%%%%%%%%%%%%%%%%%%%%%%%%%%%%%%%%%%%%%%%%%%%%%%

\end{document}